# An approach to evaluation of common DNS misconfigurations

Petar D. Bojović[1], Slavko Gajin[2],
[1]Faculty of Computer Science, Union university Belgrade, Serbia
[2]School of Electrical Engineering, University of Belgrade, Serbia
*petar.bojovic@paxy.in.rs*, *slavko.gajin@rcub.bg.ac.rs*

*Abstract* - DNS is a basic Internet service which almost all other user services depend on. However, what has been perceived in practice are a lot of inconsistencies and errors in the configuration of servers that cause different problems. The majority of such cases are included in this research with the aim of identifying and classifying the major problems of DNS availability, performance and security. In order to analyze these problems in correlation with DNS administrators working practice, we have developed a methodology and tool for testing, quantifying and analysis of DNS misconfigurations. The methodology and tool were applied on three heterogeneous domain categories – the most popular Internet domains, academic domains and one national top level domain. Our results confirm relatively high percentage of misconfigured domains, especially in the academic and national categories. However, we have shown that fixing the configuration on relatively small number of name servers can have significant impact to great number of domains. Proper domain management, permanent testing and collaboration with other administrators are identified as measures to improve domains operation, stability and security.

## I. INTRODUCTION

DNS is a global scale hierarchical service, which primary role is to resolve domain names on the Internet in order to facilitate easier mapping to device IP addresses across the network [1]. Due to the need for service high availability, DNS system is managed in a fully distributed and redundant manner, by configuring one or more so-called secondary servers, which contain copies of the domain definitions (zone files) from the primary server. Both primary and secondary name servers are authoritative and completely equivalent for the resolution of DNS for the same domain.

Though defined back in 1983 [2], there are still certain problems in the correct application of servers [1]. Since the changes in computer networks are very dynamic, DNS configurations need a constant maintenance and frequent update: new names and IP addresses are added or removed, servers are added or changed, new subdomains are opened, etc. For that reason DNS service is configured a distributed system which requires coordinated work of many administrators who are often physically separate and belong to different institutions or organizational domains. Therefore, each change can lead to potential errors, especially those changes which need to be done simultaneously in multiple domains, such as configuration of secondary servers or servers on the parent level.

Some problems with DNS configuration can easily be noticed because they influence the availability of the application service (e.g. server non-authoritativeness), others influence efficiency, reliability or domain security (e.g. Public Zone Transfer), while some problems only occur occasionally or stay hidden until more problematic errors appear.

Our motivation to extend the research in this area was to raise awareness about DNS configuration importance and help DNS administrators to better manage their domains. The objective was to identify the most frequent DNS misconfigurations, analyze their implications to availability, performance and security in correlation to working practice of DNS administrators. We have contributed to this goal by providing methodology and tool for testing, quantifying and analysis of DNS misconfigurations.

The rest of the paper is organized as follows. Section 2 makes and overview of published related work. Section 3 defines the methodology used for single domain testing. Section 4 presents the results of the massive testing of domains in three chosen datasets. Section 5 concludes our research and underlines the directions on our further work.

## II. RELATED WORK

Few studies have so far dealt with the potential causes of DNS problems and domain testing with the aim of acquiring statistical data to indicate the real-life functionality [3]. DNS have been tested by clients on different physical locations and, according to the gathered data, 26% of the names are defined in the CNAME record [4]. Therefore, the longest chain of their resolution can be as much as 4 steps long. Research has been done on DNS problem causes and they have been classified (lame delegation, reduced redundancy, Cyclic Zone Dependency) [5][6]. The problem of DNS security which occurs due to an incomplete Glue record has also been pointed out [7].

There are project that have tested DNS capability on large number of domains from client side [7], or ISP side [8]. Results show tested DNS capability like Glue record, Mixed-case, EDNS, DNSSEC, NXDOMAIN, recursion, port prediction, IPv6, SPF, DKIM.

In the paper [9] authors have combined different DNS capability tests like Public Zone Transfer (PZT), DNSSEC, locality, etc, and presented statistics for large number of domains (about 60% of all domains).

Certain papers have also made a contribution by quantifying domains response and presenting it as metric

[6]. They have defined MSQ (Minimal number of server queries) metric and conclude that average MSQ is 3.48 and about 60% of tested domains had 3 or less MSQ.

Most of the related researches consider partial issues of DNS functionality, performed in different time during the last decade. They are pointing out a significant DNS misconfiguration, but there were no comprehensive cause analyses of these problems. Having in mind rapid increase of domains on the Internet (from 21 thousand in January 1993 [10] to 270 million in November 2013 [11]) occasional testing DNS configuration on the Internet domains is still required.

Our research aims to fill this gap by defining a methodology for comprehensive domain testing, metric to quantify quality of individual domains, and analysis which was demonstrated on three different heterogenic domain datasets.

## III. DOMAIN TESTING METHODOLOGY

### A. Single domain analysis

In the first phase of our research, a web based java application for single domain analysis has been developed using open-source project dnsjava [12]. The application tests the given domain, starting from all root name servers and tracing down all the paths throughout DNS hierarchy, ending with the servers authoritative for the domain. During this process all relevant DNS parameters from each server on the paths are collected and tested regarding compliance to the standards, recommendations and best practice. The results are systematically presented using an interactive web based graphical interface, pointing out errors, warning and notifications. The application is publicly available as an on-line service for domain testing under the name NetVizura DNS Checker [13].

### B. Domain tests

Based on the insight into testing results of many public domains obtained using the application described above, and by analyzing functional characteristics of the DNS service, the most important tests which indicate DNS service malfunctioning, have been identified as follows:

- *Name server is unavailable via UDP* - Domain has correctly registered authoritative name server both on parent level and domain level itself, but at least one domain name server is unavailable via UDP protocol. If the server is selected for the name resolution, this malfunction results to time out period, which slows down the system response from the user perspective. Which server is going to be included in the query resolution process is out of the user control. Therefore, the system response is unpredictable and it takes from a few milliseconds in a regular case to a few seconds when the unavailable server is selected.
- *Name server is unavailable via TCP* - Domain has at least one name server which is unavailable via TCP protocol for the user resolutions. TCP protocol is considered as required part of full DNS protocol implementation [14]. Its implementation is necessary when the response exceeds 512-byte limit, which is the case with zone transfer, when the number of servers (NS records) is higher than 6, or with DNSSEC responses. Otherwise, lack of TCP support does not result to any problem, and it is not considered as an error.
- *Only one authoritative name server* – Domain has only one authoritative name server, and therefore it is exposed to a single point of failure problem. Even though domain registration process usually requires at least two different authoritative name servers, it happens often that these two servers are paired via CNAME record, both referring to the same IP address. Another scenario which leads to this problem is when the secondary name server never gets configured as a secondary server, or it stops working as a secondary. Consequently, when this name server is involved in the resolution, it sends non-authoritative response which results to slower resolution.
- *Non-authoritative name servers defined on the parent level* – Domain has correctly defined name server on the parent level, but it is not authoritative for the domain. From the user perspective, this problem leads to timeout, redirection and repeated query which slows down the resolution process and degrades the user experience. However, the probability of an unsuccessful query in this case depends on the total number of servers defined on the parent level for the given domain. Having small number of name servers defined at the parent level (2 or 3) gives results to greater probability of failed queries (50% or 33%).
- *"Stealth" name servers* – Domain has an authoritative name server defined on the domain zone, but it is not registered on any parent server. Even though this name server functions properly, having parent servers unaware of its existence, it is therefore "stealth" for the rest of DNS service and useless for the name resolution.
- *Loops in the resolution process* - Resolution loop can occur when name server on the parent level redirect request to the domain name server which appear to be non-authoritative for the domain, and then return reference back to the parent domain. The resolution process is slower down, but users and domain administrators generally are not aware of the problem.
- *Enabled Public Zone Transfer* - Public Zone Transfer means that everyone is allowed to access and copy whole domain content from name server, which potentially enables hackers to conduct Reconnaissance attacks [2]. Although the majority of authoritative name servers deny the Public Zone Transfer, only one name server with this option enabled is enough to put the whole domain security in danger.
- *Enabled Public Recursive* – When a name server supports public Recursive Resolution, any user can use it for DNS resolution. It exposes the server to DDoS attack and increases the risk of the domain functioning being disrupted. These servers may be used to conduct DDoS attacks to other name servers since a single request by client can generate multiple requests from this name server to other name servers [15].
- *Synchronization with the secondary servers* – Synchronization problem occur when some secondary

name servers failed to download data from the primary name server. This problem can result to resolution of inconsistent DNS data.

- *Close physical locations of name servers* – The requirement of multiple name server per zone is made for purpose of redundancy. However, it is not rare that all name servers are located at the same local network. Paper [16] describes that domains with week-spots at Dependency Graphs can be easily jeopardized with attack on small number of servers. In this case the domain is faced with a potential single point of failure (the local network), which increases the risk of service malfunction.
- *Reverse mapping* - PTR record enables reverse resolving IP address to symbolic names. Many security aspects, like SPF (Sender Policy Framework) [17], today lean on forward/reverse technique for checking the correctness of DNS records, which can still cause problems on a large number of domains.
- *IPv6 support* – The support for IPv6 is tested by checking AAAA record of NS, MX and WWW records. Defining AAAA records for name servers which do not have the IPv6 support slows down query resolution because the request for resolution is first sent via IPv6 protocol, and only when an error occurs, it is sent via IPv4 protocol (Dual Stack solution for the transition to IPv6).
- *DNSSEC support* – Support for secured DNS protocol prevents some of integrity attack on DNS. In this case each domain is tested for protected NS and A records. However, in full DNSSSEC support all zones from the root to the bottom have to support and implement DNSSEC [18].

### C. Domain state quantifying metric

To quantify domain state, as a consequence of DNS misconfiguration, the results obtained from the tests described in the text above, need to be transformed into a single metric.

Firstly, testing domain $d$ with the test $i$ gives results $T_i(d)$, defined as follows:

$$T_i(d) = \begin{cases} 0, & \text{test has passed (no error)} \\ 1, & \text{test has failed (error state)} \end{cases} \quad (1)$$

However, these testing indicators have different impact to domain functionality. Some problems affect basic functionality and they are critical to domain operation, such as problems with authoritativeness, while others influence reliability, leading to a single point of failure, or efficiency which results to unstable service response. Public Zone Transfer and Public Recursion expose domains to serious security vulnerabilities. Other tests do not necessarily imply malfunction or misconfigurations, but rather indicate the state of domain with regard to new technology penetration (IPv6 and DNSSEC) or reflect DNS administrators' principle that can be included in the metric with minor impact.

Quantifying the state of any domain therefore assumes the *Weight factor* which reflects the relevance of test $i$, denoted as $W_i$. We have chosen the scale from 0 to 10, where greater Weight factor means more problematic state of the domain, while 0 means that the test is effectively skipped from the metric. Values of the Weight factors chosen in this research are given in the Table 1.

Additionally, there are some tests mostly related to name servers operation, whose impact to the domain functionality directly depends on the number of misconfigured name servers. As an example, if one of five name servers is unavailable via UDP, the problem is manifested in 20% cases. In general, if $N_{err,i}(d)$ of $N_{tot,i}(d)$ name servers for domain $d$ are in error state against test $i$, the probability of the problem appearance is:

$$P_i(d) = \frac{N_{err,i}(d)}{N_{tot,i}(d)} \quad (2)$$

TABLE 1. WEIGHT FACTORS AND SCALED FACTORS

| No. | Test | Weight Factor | Scaled Factor |
|---|---|---|---|
| 1 | Unavailability via UDP | 10 | S |
| 2 | Unavailability via TCP | 4 | S |
| 3 | Domains with only one authoritative DNS server | 8 | 1 |
| 4 | Non-authoritative server defined on the parent level | 8 | S |
| 5 | "Stealth" DNS servers | 4 | S |
| 6 | Loops in the resolution process | 6 | 1 |
| 7 | Enabled Public Zone Transfer | 5 | 1 |
| 8 | Enabled public recursive resolution | 5 | S |
| 9 | Synchronization with the secondary servers | 2 | 1 |
| 10 | Close physical locations of DNS servers | 2 | 1 |
| 11 | Incorrect reverse mapping (PTR record) | 2 | 1 |
| 12 | IPv6 support | 2 | 1 |
| 13 | DNSSEC support | 2 | 1 |

In most cases, there are 2 or 3 name servers per domain [9][19] and typically only one misconfigured server when the problem exists, which gives 50% or 33% of the problem appearances respectively. Obviously, these Weight factors need to be scaled with the probability of problem appearance from the user perspective. If we dimension the maximum Weight factor for the case where the error appears with 50% probability (1 of 2 name servers is misconfigured) or even higher (3 of 4 name servers are misconfigured, which is possible but very rare case), we can involve the *Scaled factor* for test $i$ and domain $d$, denoted $S_i(d)$, as follows:

$$S_i(d) = \max\left(1, \frac{P_i(d)}{50\%}\right) \quad (3)$$

Tests which involve the Scaled factor are marked in Table 1 with label *S* in the last column, while other tests have Scaled factor of 1.

Finally, the total quantifying metric for domain $d$, denoted $M(d)$, is sum of the Weight factors scaled by the Scaled factor for each test which returns error state, given by the following equation:

$$M(d) = \sum_i (W_i * S_i(d) * T_i(d)) \quad (4)$$

In order to have the metric in the chosen range from 0 to 10, it needs to be normalized by the maximum measured metric ($M_{max}$) and the maximum chosen value (10):

$$M_N(d) = 10 \frac{M(d)}{M_{max}} \quad (5)$$

where maximum metric is given by:

$$M_{max} = \max(M(d)) \quad (6)$$

*D. Domain testing tool*

Based on the testing methodology and quantifying metric, we have developed the application for massive domain testing, in order to analyse operation, performances and security issues of huge number of domains. Besides the module for testing individual domains, the application also includes the following modules: scheduler for domain tests, central database, portal for domain organization and result displaying, and module for the analysis of test results.

The scheduler takes domains from the central database, testing it using the basic testing module and saves the results into the database. The testing intensity was dimensioned not to be too "aggressive" for individual servers. Therefore the module for scheduling is configured to test one domain per second, randomly choosing the order of domains for testing.

The analysis module prepares analytical reports of testing results from the database, while the web portal allows domain management, data organization and the result reporting.

*E. Domain datasets*

In order to measure DNS misconfiguration and correlate it to the working practice of DNS administrators, we have chosen three very different groups of domains, i.e. datasets.

The first dataset includes domains of research and education institutions which belong to European National Research and Education Networks (NREN). Total of 11,263 domains were gathered from 30 NRENs willing to participate in the research. The academic community is traditionally characterized by an open collaboration among the staff, both internally within the country or internationally between NRENs. In this dataset we expected to find more distributed DNS structure with seamless collaboration of DNS administrators.

The second dataset includes all domains of ".rs" ccTLD, which belongs to Republic of Serbia. Serbia is a country with middle level of Internet penetration were about 53% of households have the Internet connection in 2013 [20]. The dataset with total of 57,004 domains was provided by the Serbian National Internet Domain Registry - RNIDS, which partially sponsored this research. They belong to wide range of customers, such as commercial companies, non-profit organizations, or individuals. Domains are registered and usually hosted by the authorized registers, which are mostly ISP companies. They are competitors on the market with less willingness to collaborate, even on the technical matters.

The third dataset consists of 1,000 the most popular domains, according to [21]. Majority of them are domains under ".com" domain and many ccTLD (64.3% .com, 5.1% .net, 1.8% .org, and other different ccTLD). Since these are popular and frequently used domains it is expected that they are supported by more professional DNS administration.

The purpose of massive domain testing and analysis is to compare DNS misconfiguration within these three different categories of domains, labelled as NREN, ccTLD and Top in the rest of the text.

IV. RESULTS AND ANALYSIS

The domain testing and analysis of the results were performed on October, 2014. The testing of a total number of 79,933 domains lasted for about 40 hours. During this process, a certain number of domains were unavailable due to the impossibility of finding the authoritative servers related to these domains (6.44% of the NREN domains, 17.41% of the ccTLD domains and 0.4% of Top domains). These were either incorrectly registered domains or domains whose subscription had expired, and therefore were excluded from the analysis.

The rest of the section presents summary results of tests performed on the most important indicators, stated in chapter 3, simultaneously for three different domain datasets.

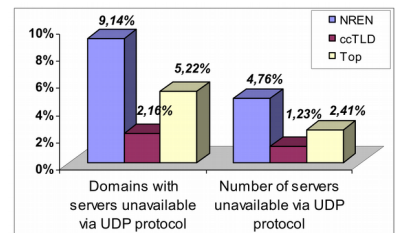

Fig. 1. Problem of unavailability via UDP protocol

*A. Unavailable via UDP protocol*

Around 9% of NREN domains, 5.2% of Top and, 2% of ccTLD domains have at least one server which does not respond to UDP queries, as shown in Fig. 1. Comparing to all existing name servers in these three groups about 4.76% of NREN, 1.23% of ccTLD and 2.41% of Top name servers were unavailable via UDP protocol. The reasons for domain unavailability via UDP protocol may be temporary (e.g. a server or a network are not functioning at the moment), or permanent (e.g. a badly configured name or address of a name server on the parent level). In any case, those name servers cannot respond to any DNS request.

*B. Unavailable via TCP protocol*

The results gathered by testing DNS availability via TCP protocol show that around 28.6% of ccTLD, 15.9% of NREN and nearly 29% od Top servers do not respond to TCP queries (Fig. 2).

The most frequent causes of DNS service unavailability via TCP protocol occur when a network firewall or the server itself explicitly denies TCP queries, allowing only TCP communication with the secondary servers to perform synchronization.

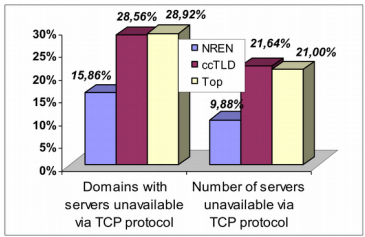
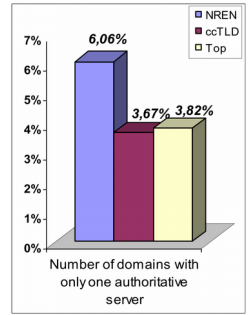
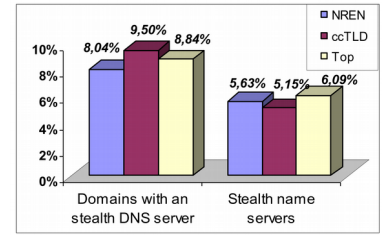

Fig. 2. Problem of unavailability via TCP protocol

Fig. 3. Domains with only one authoritative server

Fig 5. Problem of "stealth" Servers

*C.*

Domains with only one authoritative serverAlthough minimal number of authoritative name servers for any domain should be two, there is a significant number of domains in all datasets that have only one authoritative name server: more then 6% of NREN, 3.67% of ccTLD, and 3.82% of Top domains (Fig. 3.).

Our research shows that 50% of NREN domains, 70.18% of Top and only 31.97% of ccTLD domains have more than 2 name servers, while 57.93% Top, 18.58% NREN and 16.42% of ccTLD domains have more than three servers. It is pointed out in the research from 2009 [5] that 65% of domains have only two authoritative servers, while 20% have three or more servers. Only Top domains, which are mostly business driven, are above that average.

*D. Domains with non-authoritative server defined on the parent level*

Fig. 4 shows that almost 5% of NREN domains have at lease one non-authoritative name servers defined on the parent domain, which are significantly higher results comparing to 1.1% of ccTLD and 1.2% of Top domains.

Taking into account the number of name servers for a domain, the probability of unsuccessful query, which is sent to non-authoritative server, is about 40%. In that case the query has to be repeated to another server.

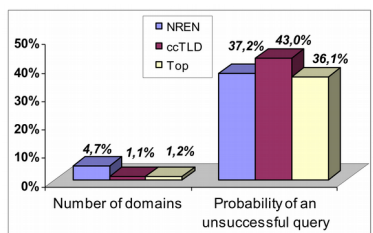

Fig. 4. Problem of non-authoritative servers defined on the parent level

*E. Domains with "Stealth" Servers*

According to Fig. 5, the problem of stealth servers is equally presented in all datasets - over 8% of domains. Those servers are authoritive but not used in the resolution process.

This indicates that domain administrators are not aware of the problem, which increases the risk of the domain failure. Therefore, it is necessary for domain administrators to correctly register name servers on the parent domain. This also demonstrates the need for much better collaboration between DNS administrators, especially with the administrators on the parent level.

*F. Domains with loops in the resolution process*

Loops in the resolution process were detected in 2.63% of NREN domains, 0.44% of ccTLD and 0.7% of Top domains (Fig. 6). Even though we are dealing with a rather small percentage of domains with detected loop error, these errors considerably slow down the resolution process and they might make the domain unavailable due to the timeout period.

The solution to this problem can be found in increasing the responsiveness of domain administrators so that a non-authoritative server wouldn't be defined in the parent domain.

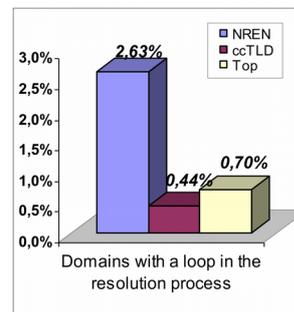
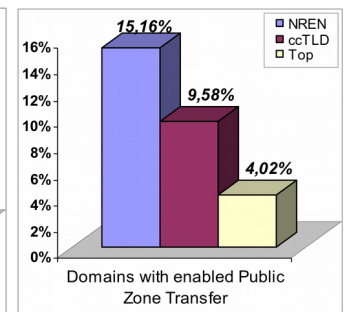

Fig. 6. Problem of domains with a loop in the resolution process

Fig. 7. Problem of enabled Public Zone Transfer

*G.*

Enabled Public Zone TransferIt is alarming that there are 15% of NREN domains, 9.6% of all ccTLD domains, and 4% of Top domains have Public Zone Transfer enabled on at least one name server (Fig. 7).

Initial testing, conducted in 2012, showed that about 22.7% domains in ccTLD dataset had PZT enabled on at least one server. In 2012, we have established communication with ccTLD. During 2013, Serbian National Internet Domain Register - RNDIS [23] have managed to significantly improve state of ccTLD domains using results of our research.

*H. Enabled Public Recursive*

Fig. 8 shows that almost 6% of NREN domains, 1% of ccTLD domains and 2.2% of Top domains have at least one server which enables Public Recursive.

Comparing to the initial testing of ccTLD domains from 2012, where this type of error was presented in 12.8% of all national domains, the latest result clearly demonstrate a significant improvement that can be done by the coordinated and collaborative DNS administration.

*I. Problem with the synchronization of secondary servers*

During the testing, a relatively small number of domains

(less than 1%) had a problem with the synchronization between the primary and secondary servers, which is shown in Fig. 9.

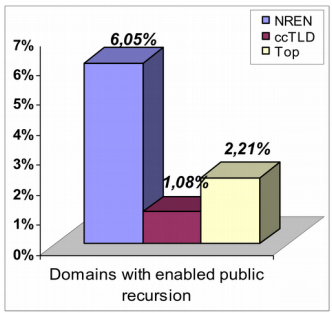
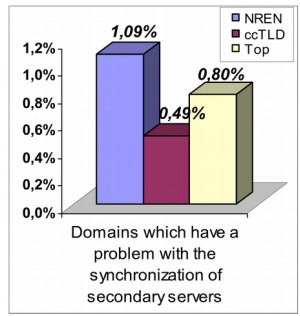

Fig. 8. Problem of public recursion

Fig. 9. Problem with the synchronization of secondary servers

*J. Problem of close physical locations of servers*

The results shown in Fig. 10 indicate that 50.44% of ccTLD domains most probably have closely located name servers, which is detected by the same IP network with a 24-bit mask. The reason for such a great percentage lays in the fact that ccTLD domains are usually hosted by authorized local registers or Internet providers. It also indicates the need for a better cooperation between local Internet providers and authorized registers in the mutual distribution of hosting services.

The situation is much better in NREN dataset (18.25%), which indicates that academic institutions are ready to provide DNS hosting to each other.

Top domains in just about 5.95% of cases have name servers in same network.

Increasing number of name servers located in different ISPs, physical network, counties or continents, decrease chance of name service unavailability and prevent single point of failure.

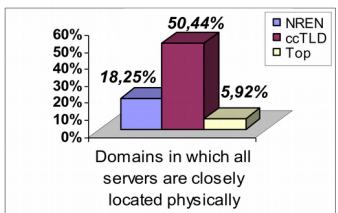
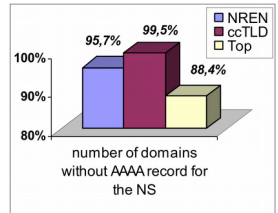

Fig. 10. Problem of close physical locations of server

Fig 11. IPv6 support

*K. Incorrect reverse mapping (PTR record)*

Reverse DNS records were tested for all unique name servers. The results shown in Fig. 12 present that 41,7% of NREN domains, 39,7% of ccTLD and over 45% of Top domains do not have PTR record at all. For those domains that have reverse record 5.3% of NREN servers, 2.8% of ccTLD, and more then 20% of Top servers are unavailable via their names (they are referred to non-existing domains). Around 4.8% of NREN, 3.8% ccTLD and 7.9% of Top domains servers have a reverse DNS record which do not match with a forward DNS record (e.g. they use alias (CNAME) records, or many domain names are being resolved to the same IP address of the Server). The consistent reverse mapping achieved at 48.2% of NREN, 53.7% of ccTLD servers, and just 26% of Top domains.

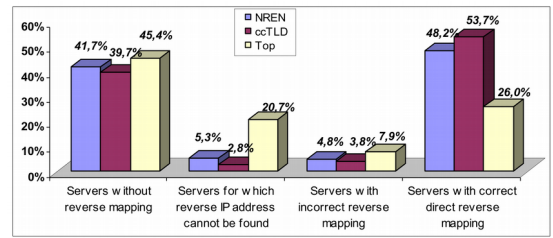

Fig. 12. Problem of reverse mapping

*L. IPv6 support*

It is worrying that a large number of domains do not have the IPv6 support in the form of adjusted AAAA records, despite the fact that DNS configuration may easily be adapted to IPv6 protocol. This case is especially seen in the tested ccTLD which is, comparing to other countries, late with the implementation of IPv6 protocol (Fig. 11).

The biggest penetration of IPv6 protocol is with most popular sites. It is slightly comparable to Project 6lab (results from 2014.) which presents that about 28% domains support IPv6 protocol [22]. That result includes 27.84% of domains with Dual Stack (IPv4 and IPv6) support and only 0.00053% Native IPv6 support.

*M. DNSSEC support*

The testing results show that the implementation of DNSSEC technology for the DNS service protection has not gone very far which is depicted in Fig. 13. The tested ccTLD still does not have the DNSSEC support on its domains, thus the entire implementation on sub-domains is not possible. Among NREN domains, only 7.9% of parent domains support DNSSEC and have made room for the protocol support on sub-domains. About 2.4% of Top domains have implemented from top to bottom DNSSEC protection.

Independently of the support on parent domains (which is mandatory for the complete DNSSEC implementation) around 1,2% of NREN domains and 11,7% of ccTLD, and just about 1% of Top domains have servers with configured DNSSEC support for the protection of NS domain records. These domains are ready for the completion of DNSSEC implementation when the parent domain enables DNSSEC protection. A certain number of servers with DNSSEC support already support additional records, such as A, AAAA and MX.

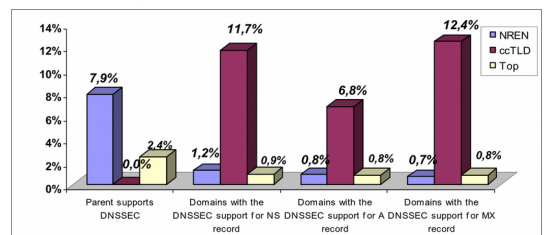

Fig. 13. DNSSEC support

*N. Normalized domain metric distribution*

Normalized domain metric is a useful technique to measure the operational state of a single domain. To have an overview of operational state of all domains in the datasets, a cumulative density function (CDF) of normalized domain metric is shown in Fig. 14. CDF graph shows percentage of domains in the datasets which have equal or smaller normalized metric that a certain value. It allows efficient comparison of tested results for different datasets.

Metric distribution graph shows that less than 1% of domains in NREN and Top dataset have passed all of our tests, while none of ccTLD domains has passed (due to lack of IPv6 and DNSSEC implementation on ccTLD level). More than half of all domains have metric less then 1 (50.1% NREN, 39.8% ccTLD, 58.9% Top). All tree dataset have about 90% of domains metric less then 2 (85.9% NREN, 89.3% ccTLD, 92.1% Top). Difference is notable up to normalized metric of 3, reaching 97% of domains in all categories.

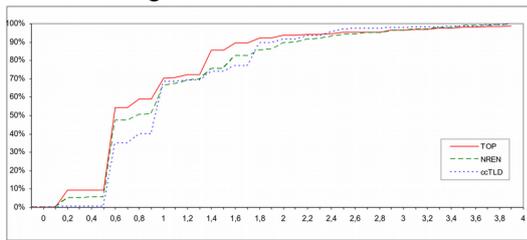

Fig. 14. Normalized domain metric distribution (CDF)

*O. Name server distribution analysis*

Since the problems of domain operation considered in this research are caused by the misconfiguration of the name servers, we have further analyzed the number of domains affected by individual misconfigured name servers for the specific problem. The purpose of this analysis is to check how many domains with certain error can be fixed by correcting configuration on relatively small number of name servers. The candidates are those name servers which host large number of domains.

Distribution of all name servers which are non-authoritative but defined on the parent zone is shown in Fig. 15. The result for ccTLD domains depict that there is a single name server involved in 6% problems of this type. It means that correcting the issue of non-authoritativeness with the parent zone for domains on only one name server, by removing server from the parent zone or reconfiguring it to become authoritative for the domains, this type of error can be significantly reduced. Furthermore, resolving this issue on top 5 name servers in ccTLD dataset, the problem would be reduced by 24.5%. This is the consequence of hosting a large number of domains on the name server which are actually misconfigured.

Greater diversity of name servers for domains in NREN and Top datasets results to lower concentration of this problem per servers, but considerable improvement can also be achieved on just a few servers. For NREN domains, resolving this error on only one server can reduce the problem by 7.7%, while fixing problem on 5 servers would resolve 14.8% domains with this issue. In Top domain dataset, resolving this issue on top 5 servers would reduce number of domains with this problem by 4.84%.

The similar misconfiguration with the parent level is the problem with "stealth" name servers. Fig. 16 shows that fixing top 5 Stealth name servers for ccTLD domains would reduce the problem by 27.5%. In NREN and Top datasets, resolving Stealth issue on top 5 servers would reduce this issue by 12% and 7.5% respectively.

Public Zone Transfer and Public Recursion are the security problems which are caused only by local misconfiguration on the name server and therefore can be more easily fixed by proper DNS administration. At ccTLD dataset Public Zone Transfer on only one name server is related to 13% of domain with this error. Disabling Public zone transfer on 5 top name servers would resolve this security issue for 35.34% affected domains. In NREN and Top domain datasets, top 5 name servers cover 15.48% and 3.7% domains exposed to this security issue (Fig. 17).

When it comes to Public Recursion, disabling this issue on only five name servers for domains in ccTLD dataset would decrease the problem by 11.18%, as it is shown in Fig. 18. Disabling Public Recursion on top 5 name servers in NREN and Top domain dataset can improve security issue of 18% and 3% affected domains respectively.

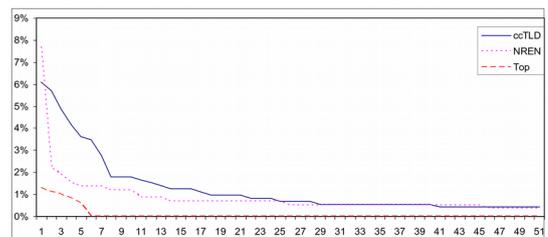

Fig. 15. Server distribution for Not-authoritative but defined on parent

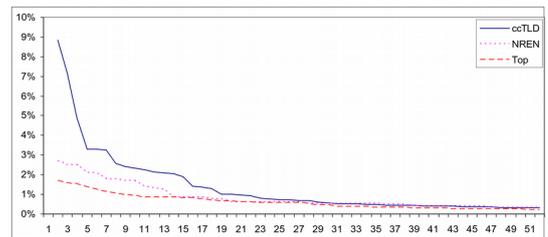

Fig. 16. Server distribution for Stealth issue

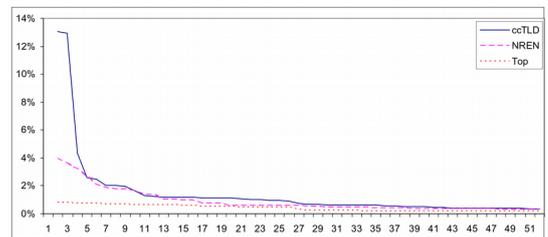

Fig. 17. Server distribution for Public Zone Transfer

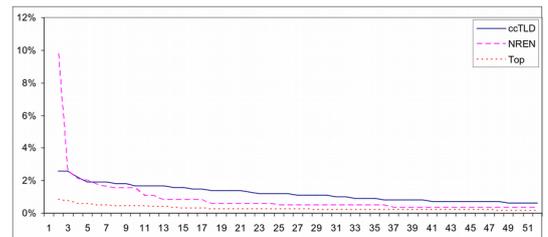

Fig. 18. Server distribution for Public Recursion issue

V. CONCLUSION

By analyzing individual problems occurring in DNS configurations we have identified the most frequent problems and classified them into 13 types of tests to indicate DNS proper functionality. Based on these indicators we have defined an efficient methodology to measure, monitor, compare and analyse domains' state. The methodology is also simple and flexible, which gives the possibility to adjust the importance of measured indicators changing the Weight factors according to the domain purpose. For example, if a domain security is an issue of higher importance, the Weight factor of Public Zone Transfer and Public Recursion, which affect a security, can be increased.

Similar to previous research reported in the literature, our work confirms that DNS misconfiguration is still an open issue. There are a considerable number of domains which have some problem with the name server malfunction (25.2% of NREN, 22.54% of ccTLD and 17.7% of Top domains have problem with either none or one authoritive server, loop, or non authoritive server on parent, or non accessible servers via UDP and TCP ports). These domains show a lower resolution performance and threatened service reliability.

Along with the proper domain administration on the primary name server, the collaboration with other DNS administrators is necessarily for correct domain functioning and operation, what appears to be the weakest point in domain maintenance. As an example, it is useless to deny Public Zone Transfer on the primary name servers as long as there is only one secondary name server which allows it. Despite the fact that it is relatively easy to check and fix PZT and Public Recursion issue problem (even by email correspondence), 18.6% of NREN, 10.2% of ccTLD and 6% of Top domains are exposed to this security vulnerabilities.

But the most critical problems are related to the authority of name servers and registration on the parent level. Therefore, a coordinated and synchronized work with the parent level is of special importance. If this collaboration failed during domain registration or on any changes during domain management, the resolution process will lead to non-authoritative servers, or contrary, some authoritative name servers can stay hidden, i.e. "stealth".

Although the collaboration in the academic community is a widely accepted practice, a large percentage of errors show that it does not always follow a dynamic nature of domain changes. Many domains in ccTLD category are also suffer from misconfiguration, but still less than those in academic community. Significantly lower errors with the most popular domains ensure that DNS management is a professional business driven process, what is the case in commercial sector.

The analysis of name servers involved in the individual problems shows that misconfiguration on one name server most probably affects all hosted domains. Consequently, we have shown that the problems can be significantly minimized by fixing the configuration on relatively small number of name servers. This is especially remarkable when large number of domains is concentrated on several name servers, what is the case with the local registers and ISP companies within ccTLD.

And finally, in addition to careful technical domain maintenance and synchronized work with other administrators, a permanent monitoring and testing is of outmost importance. We contributed to this task by providing an on-line service for individual domain testing [13], which offers an opportunity to check, analyse and isolate DNS misconfiguration problems.

The further research will focus on permanent DNS monitoring system and its impact to reduction of DNS misconfiguration over the time.

ACKNOWLEDGMENT

The research presented in this paper was partly funded by Serbian National Internet Domain Registry (RNIDS) [23]. The authors would also like to thank RNIDS and European NRENs who have provided domains to complete the testing datasets.